\RequirePackage{xcolor}
\documentclass[lettersize,journal]{IEEEtran}
\usepackage{cite}
\usepackage{float}
\usepackage{amsmath,amssymb,amsfonts}
\usepackage{algorithmic}
\usepackage{graphicx}
\usepackage{textcomp}
\usepackage{multirow, makecell}
\usepackage{booktabs}
\usepackage{tabularx}
\usepackage{chemformula}
\usepackage{ragged2e}
\usepackage[affil-it]{authblk}
\usepackage{epstopdf}
\usepackage{pifont}
\newcommand{\cmark}{\ding{51}}%
\newcommand{\xmark}{\ding{55}}%

\begingroup\expandafter\expandafter\expandafter\endgroup
\expandafter\ifx\csname IncludeInRelease\endcsname\relax
  \usepackage{fixltx2e}
\fi
\def\BibTeX{{\rm B\kern-.05em{\sc i\kern-.025em b}\kern-.08em
    T\kern-.1667em\lower.7ex\hbox{E}\kern-.125emX}}
\begin{document}
\title{RRWaveNet: A Compact End-to-End Multi-Scale Residual CNN for Robust PPG Respiratory Rate Estimation}



\author{Pongpanut~Osathitporn$^{\dag}$, Guntitat~Sawadwuthikul$^{\dag}$,  Punnawish~Thuwajit$^{*}$,  Kawisara~Ueafuea, Thee~Mateepithaktham, Narin~Kunaseth, Tanut~Choksatchawathi, Proadpran~Punyabukkana$^{*}$, Emmanuel~Mignot and Theerawit~Wilaiprasitporn$^{*}$, \IEEEmembership{Senior Member, IEEE}
\thanks{This work was supported by PTT Public Company Limited, The SCB Public Company Limited, Office of National Higher Education Science Research and Innovation Policy
Council (C10F630057) and National Research Council of Thailand (N35A650037) \textit{($^{\dag}$P.~Osathitporn and G.~Sawadwuthikul equally contributed to this work.) ($^{*}$co-corresponding authors: P.~Thuwajit, P.~Punyabukkana, and T.~Wilaiprasitporn).}}
\thanks{P.~Osathitporn and P.~Punyabukkana are with Department of Computer engineering, Faculty of Engineering, Chulalongkorn University, Phayathai Road, Pathumwan, Bangkok 10330, Thailand (e-mail: proadpran.p@chula.ac.th).}
\thanks{G.~Sawadwuthikul is with the School of Computing, Korea Advanced Institute of Science and Technology (KAIST), Daejeon 34341, South Korea.}
\thanks{P.~Thuwajit is with the University of Wisconsin Madison, Madison, Wisconsin, U.S. (e-mail: thuwajit@wisc.edu).}
\thanks{T.~Mateepithaktham is with Suankularb Wittayalai School, Phra Nakhon, Bangkok 10200, Thailand.}
\thanks{K.~Ueafuea, N.~Kunaseth, T.~Choksatchawathi and T.~Wilaiprasitporn are with Bio-inspired Robotics and Neural Engineering (BRAIN) Lab, School of Information Science and Technology (IST), Vidyasirimedhi Institute of Science \& Technology (VISTEC), Rayong, Thailand (e-mail: theerawit.w@vistec.ac.th).}
\thanks{E.~Mignot is with Stanford University, Palo Alto CA 94304, U.S..}
}%

\maketitle

\begin{abstract}
Respiratory rate (RR) is an important biomarker as RR changes can reflect severe medical events such as heart disease, lung disease, and sleep disorders. Unfortunately, standard manual RR counting is prone to human error and cannot be performed continuously. This study proposes a method for continuously estimating RR, RRWaveNet. The method is a compact end-to-end deep learning model which does not require feature engineering and can use low-cost raw photoplethysmography (PPG) as input signal. RRWaveNet was tested subject-independently and compared to baseline in {\color{red}four} datasets (BIDMC, CapnoBase, WESAD, and SensAI) and using three window sizes (16, 32, and 64 seconds). RRWaveNet outperformed current state-of-the-art methods with mean absolute errors at optimal window size of 1.66 ± 1.01, 1.59 ± 1.08, 1.92 ± 0.96, {\color{red}and 1.23 ± 0.61} breaths per minute for each dataset. In remote monitoring settings, such as in the {\color{red}WESAD and SensAI datasets}, we apply transfer learning to improve the performance using two other ICU datasets as pretraining datasets, reducing the MAE {\color{red}by up to 21$\%$}. This shows that this model allows accurate and practical estimation of RR on affordable and wearable devices. Our study also shows feasibility of remote RR monitoring in the context of telemedicine and at home.
\end{abstract}

\section{Introduction}
\label{sec:introduction}
\IEEEPARstart{I}{n} the hospital healthcare sector, RR, or the breathing rate, alongside pulse rate, body temperature, and blood pressure is one of four vital signs that is used to summarize clinical status for patients \cite{rrreview}. Indeed, several fatal diseases of the respiratory and cardiovascular systems are reflected by acute RR changes. 

Despite the need for continuously measuring RR in the hospital and elsewhere, an effective method to do so is not available. Rather, RR measures are most often performed by human personnel sporadically counting chest wall movements. This time-consuming procedure is labor intensive and requires trained staff \cite{inaccurate}.

Instead of performing manual measurements, recent studies \cite{rrwearable, rrecg, ppgreview} have attempted to more accurately extract RR continuously using non-invasive and cheaper approaches that take advantage of other biosignals that are easier to measure. On such signal has been the PPG, a signal commonly embedded in current wearables.  

The PPG signal is obtained by supplying a supplementary light source and detecting the amount of reflected light from blood and tissues so that changes in blood volume are measured. Clinical applications include measurements of pulse rate and of blood oxygen saturation (\ch{SpO2}), two measures that correlates with RR \cite{rrreviewecgppg}.

Pushing the limits of respiratory rate estimation performance utilizing PPG signals, our contributions in this work are listed below.
\begin{itemize}
    \item We propose \emph{RRWaveNet}, an end-to-end, compact, and explainable respiratory rate estimator using a multi-scale convolution and residual CNN to analyze the PPG in a subject-independent manner. \emph{RRWaveNet} outperforms the other four state-of-the-art approaches and shows robust performance across multiple datasets shown in Table \ref{table:1}.
    \item We introduce a transfer learning approach from dataset-to-dataset and device-to-device to enhance the model performance, especially those acquired from wearable devices in which the lack of training data presents. This strategy results in a decreased MAE by up to 21$\%$. 
\end{itemize}


Starting with Section \ref{sec:related}, this article discusses background and related work. Data preprocessing on the signals and our proposed neural architecture are elaborated in Section \ref{sec:methodology}. The demonstration and results of our experiments follow in Section \ref{sec:experiments}, where we compare the performance of our approach when tested on four benchmark datasets to other recent state-of-the-art methods. We discuss the explainability of the model and possible improvements in Section \ref{sec:discussion} and conclude the article in Section \ref{sec:conclusion}.

\section{Related Work}
\label{sec:related}

\begin{figure*}[!h]
  \center
  \includegraphics[width=0.9\textwidth]{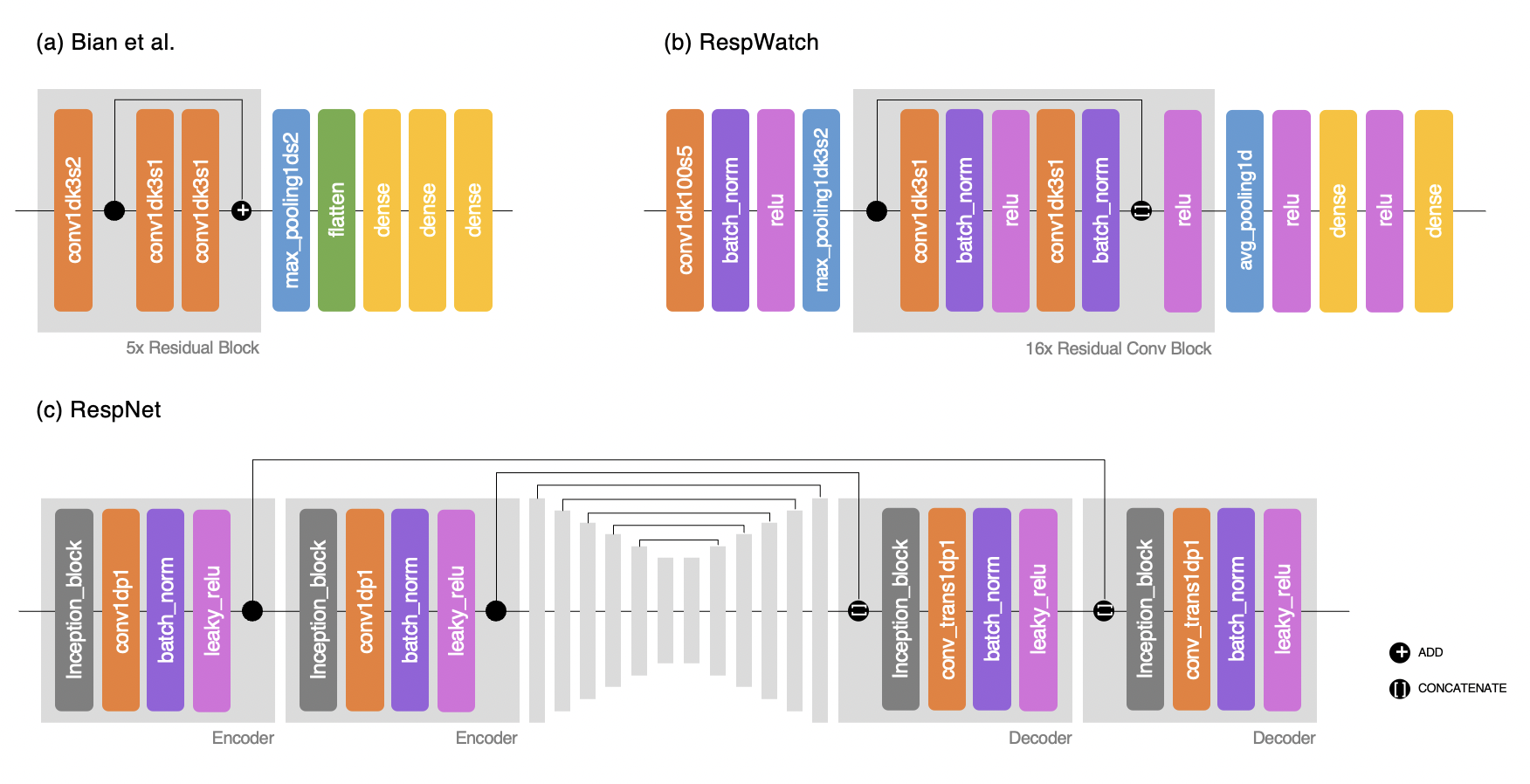}
  \caption{Deep neural network architectures for respiratory rate estimation from recent studies: (a) Bian et al. \cite{bian} applies ResNet blocks and tunes their hyperparameters using Bayesian optimization, (b) \cite{respwatch}, from Dai et al., adapts the convolutional neural network to help capture the motion artifcats in the PPG signal, and (c) Ravichandran et al. \cite{respnet} proposes an IncResU-Net-like architecture consisting of multiple encoders and decoders.}
  \label{fig:architectures}
\end{figure*}

In the last few decades, many studies have attempted using PPG signals to predict respiratory rate. We classify existing approaches into three groups: signal processing, machine learning and hybrid approach. Among the studies cited, we focus only on recent state-of-the-art models and include these as baseline models for the purpose of performance comparison in Section \ref{sec:experiments}. The remaining parts in this section overview the methodology used by others to approach this problem.

\subsubsection{Signal Processing}

For several decades, signal processing has played an important part in sensor applications. In RR estimation, autoregressive (AR) models have been most employed. Recently, Pimentel et al. \cite{Pimentel2017}  proposed a technique that involves multiple AR models. In this strategy, respiratory-induced intensity, amplitude, and frequency variation (RIIV, RIAV, and RIFV), were algorithmically extracted and normalized, and each variation signal reconstructed into multiple AR models tested with varying model orders. The frequency spectrum of each AR models derived using all possible model orders and variational signals were then compiled into a single spectrum that shows the amplitude median. Finally, the output RR was estimated by converting the frequency with the maximum amplitude into breath per minute.

Haddad et al. \cite{Haddad2021} fused the RR estimation from the same set of respiratory-induced variation signals by either choosing or averaging from those with low estimating variance. Another work from Karlen et al. \cite{Karlen2013} extracted pulses from the PPG signal by monitoring rapid change of the PPG slope and computing RR using a combination of fast Fourier transform and mean calculation from variation signals. Despite the instructive nature of these methodologies, processing was ruled based and not robust to the presence of artifacts \cite{bian, Aqajari2021}.

\subsubsection{Deep Neural Network}

More recently, the emergence of neural networks and technological advancement of computing resources have led to the development of deep learning methods that are very efficient at extracting important information and making good prediction, providing sufficient data is available. Several studies and applications of biosignals have introduced deep neural architectures to tackle the problem, especially the inclusion of convolutional and recurrent modules. Convolutional neural networks (CNN) is a subclass of deep learning architecture capable of capturing important features from 1-dimensional time series data as frequently appeared in biosignal inputs. Previously, CNN has been widely used to process health-related signals such as the electrocardiogram (ECG), electroencephalogram (EEG), as well as PPG in various applications. Complementary to CNN, residual blocks have been added that encode temporal information and prevent performance degradation \cite{Yu2022}. Promising examples include heart rate estimation \cite{Xu2020}, blood glucose monitoring \cite{Zhang2020}, and seizure detection \cite{eegwavenet}.

In RR estimation, as illustrated in Figure \ref{fig:architectures}, Ravichandran et al.\cite{respnet}  adopted IncResU-Net to reconstruct respiration signals from the PPG. Defined as RespNet, the architecture itself consists of a series of encoders and decoders built from dilated residual inception blocks and skip connections. RR estimation can be done explicitly on the RespNet’s output using a peak detector. 

{\color{red}
Aqajari et al. \cite{Aqajari2021} leveraged the reconstruction ability by introducing cycle generative adversarial networks to construct PPG and the respiration signal from one another. In addition to \cite{respnet}, they also extracted the RR from the reconstructed respiration signal, compared with the ground truth, and included the error to the optimizing function to ensure that the reconstructed signal was valuable for RR estimation. Summarized in Table \ref{table:comparison}, we considered both networks not purely end-to-end as they required another peak detector to estimate the RR from the output respiration signal. 
}

Bian et al. \cite{bian}, on the other hand, proposed an end-to-end network by utilizing ResNet blocks as the backbone architecture and attaching fully-connected layers to predict the RR. Recently, another study by Dai et al. presents RespWatch \cite{respwatch}, a lightweight end-to-end deep learning estimator that can be deployed on smart wearable devices. RespWatch extended the residual neural network and demonstrated an outstanding performance.

\subsubsection{Hybrid Approach}

In addition to these individual approaches, studies that apply both signal processing and machine learning approaches in an hybrid fashion have also been developed. Dai et al. \cite{respwatch}, for example, enhanced the performance of a deep learning estimator by Reassembling it with another signal processing estimator. They assessed the quality of the PPG signal and decided to apply signal processing estimator instead of deep learning if the quality was above a certain threshold (EQI $> 2.3$) following a grid search. Signal processing was found to be more accurate with the presence of a moderate amount of noise artifacts, while the deep learning approach, a more computation intensive methods, tolerates noise artifacts better.


\begin{figure*}[!t]
    \centering
  \includegraphics[width=0.75\textwidth]{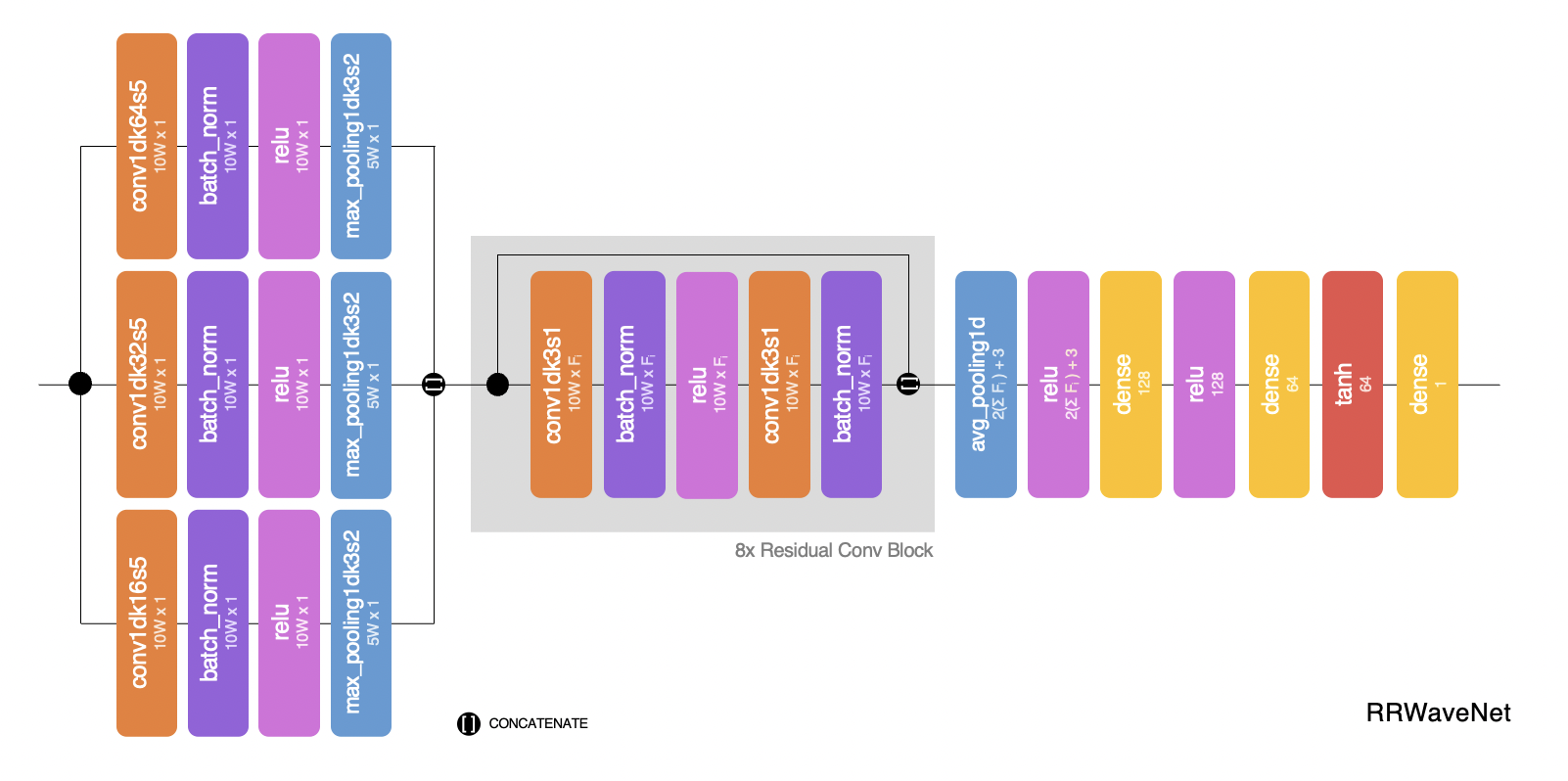}
  \caption{Our proposed architecture, \emph{RRWaveNet}. Composed of three modules, \emph{RRWaveNet} involves the multi-scale convolution (left), the deep spatial-temporal residual blocks (center), and the respiratory rate estimator (right). Each layer's title is abbreviated at the top row for simplicity and the shape of the output tensor after each layer is specified below its title. For example, conv1dk32s5, the leftmost layer in the center module, refers to a 1D-convolutional layer with a kernel of size 32 and a stride of 5, resulting a $(10W, 1)$ tensor.}
  \label{fig:rrwavenet}
\end{figure*}

\subsubsection{Transfer Learning}

Transfer learning utilizes knowledge gained from one domain of application and applies it to another, similar domain\cite{metasleeplearner}. In machine learning, it is common practice to employ pretrained weights that have been trained on a large-scale dataset and to fine-tune the model using a target dataset\cite{9522688}. This approach allows the final model to fine tune itself more rapidly to a specific application when data is sparse in the final target dataset. 

With this motivation, many studies have applied transfer learning to extract interesting outputs from the PPG because of its potential. An example  \cite{8856989} used transfer learning to estimate heart rate from PPG. To do so, these authors pretrained a model using the IEEE SPC 2015 dataset and fine-tuned it with sparser in-house datasets. Brophy et al.\cite{9156120} similarly used transfer learning to the problem of blood pressure estimation using more abundant wrist-worn sensor data to apply it in the presence of a limited dataset. Instead of time-series data, the training samples were captured as images from signal plots and trained on a network pretrained with the ImageNet dataset. \cite{9522688} on the other hand, created a feature extractor that could be multi-purposed to use either rPPG or PPG for blood pressure assessment.

As wearable devices are becoming more accessible and accurate, RR estimation is an obvious application. Since only a few existing datasets contain both PPG and gold standard respiratory rate signals, we consider using these datasets as the source domain and data collected from wearable devices as the target domain. With the potential of transfer learning, we expect to resolve two crucial concerns: data scarcity and difference in data distribution between source (collected using pulse oximeter at high quality) and target (collected using wearable devices that may contain noise) datasets. In Section \ref{sec:exp2}, we demonstrate that transfer learning using \emph{RRWaveNet} leads to a large performance improvement in our setting. 

\section{Methodology}
\label{sec:methodology}

Before PPG signals were split into training and test sets, all low-quality signals, which were assumed to be artifactual \cite{Pimentel2017} were segmented and excluded from the dataset using a procedure elaborated in Section \ref{sec:preprocess}. Our proposed model architecture is mentioned afterwards in Section  \ref{sec:rrwavenet}.

\subsection{Data Preprocessing}
\label{sec:preprocess}
\subsubsection{Signal Resampling}

All input PPG signals were resampled at a frequency rate of 50 Hz. 
{\color{red}As we wanted an abundant number of low-noise samples to train the estimator, we applied the \emph{sliding subwindow technique} \cite{respwatch} to generate more training samples -- all of which need to be qualified based on the measured SQI, explained in the following subsection.}
Similar to the convolutional kernel approach used in a neural network for image processing tasks, this technique re-crops the input signal and produces shorter sequences. The cropping kernel can be thought of 1D kernel with a length of $W$, the window size, and a stride of 2 seconds. In this work, we study the effect of the window size by varying  $W=16$, $32$, and $64$  seconds. Through this processing, we obtain signal samples of length $50W$, which is the product of the sampling rate and the window length in seconds.

\subsubsection{SQI Measurement}

In addition to subwindowing input samples, we identified and discarded low-quality signals by following guidelines reported in Pimentel’s work\cite{Pimentel2017}. Specifically, a signal quality metric (SQI), which depends on the non-flatlining points in the signal and the level of agreement between two peak detectors, was used, and samples with an SQI value less than $0.9$ were removed. The SQI metric is calculated by multiplying the two factors: $K$ and F1-score.

To determine $K$, flatlines in the windowed sample were detected and counted using hysteresis thresholding. If the change was less than $0.02$ for at least $30$ consecutive points in the time series, those points were considered as flatlining. Consequently, the measurement $K$ is calculated from
\begin{align}
    K=\frac{\text{number of non-flatlining points}}{\text{total number of points}}.
\end{align}

Another factor that contributes to SQI measurement requires two decent algorithmic peak detectors \cite{peak1, peak2}. After peak detection results from the two detectors are obtained, each pair of detected peaks is considered to be coincident if the two peaks appear less than 150 milliseconds apart. The detection results are compared by measuring their agreement using F1-score evaluation.

\begin{table}
\caption{Comparison of State-of-the-Art Approaches \\ for Respiration Rate Estimation}
\centering
\resizebox{\columnwidth}{!}{%
\begin{tabular}{lcccc}

\toprule[0.2em]

Model & End-to-End & Deep Learning & Total Parameters & Sampling Rate \\ 

\midrule[0.1em]

RespWatch \cite{respwatch} & \cmark & \cmark & 23,549,416 & 50 \\
Deep Learning \cite{bian} & \cmark & \cmark & 117,793 & 50 \\
RespNet \cite{respnet} & \xmark & \cmark & 5,446,279 & 256 \\
{\color{red}CycleGAN \cite{Aqajari2021}} & {\color{red}\xmark} & {\color{red}\cmark} & {\color{red}17,771,304} & {\color{red}30} \\
Autoregressive \cite{Pimentel2017} & \xmark & \xmark & 0 & 300/125/50 \\
RRWaveNet (ours) & \cmark & \cmark & 6,857,594 & 50 \\

\bottomrule[0.2em]
\end{tabular}}
\label{table:comparison}
\end{table}

\subsection{RRWaveNet}
\label{sec:rrwavenet}

Our proposed end-to-end deep neural network architecture, \emph{RRWaveNet}, consists of three modules: a multi-scale convolution, deep spatial-temporal residual blocks, and a respiratory rate estimator. The model is illustrated in Figure \ref{fig:rrwavenet}, and detailed below.

\subsubsection{Multi-Scale Convolution}

Inspired by the multi-scale convolution used in the grid modules of various inception networks\cite{inception}, we employed a multi-scale convolution module to capture features of the signal at different resolutions. Using three branches in parallel, input was passed through three channel-wise convolutional layers before being concatenated at the end. The three convolutional layers contain different filter sizes, learning the appropriate weights for each convolutional resolution (and in turn, the frequency of the recurring features in the input signal) independently before joining the acquired information for following modules. Our inclusion of a multi-scale convolution module resulted in more robust feature extraction, {\color{red}as we compared explicitly in Section \ref{sec:discussion}}, allowing an overall smaller architecture described as follow.

Each branch accepts a common 1D-input of shape $(S_RW, 1)$ where $W$ refers to the same window size that has been selected during preprocessing with the sliding window technique and $S_R$ corresponds to the sampling rate, which, in this article, equals $50$ as we mentioned in Section \ref{sec:methodology}. The kernel size of the convolutional layer in each branch is specified in Figure \ref{fig:rrwavenet}. Followed by a batch normalization layer, a ReLU activation layer, and a max-pooling layer, each branch outputs an $(\frac{S_RW}{5}, 1)$ tensor. Three tensors of the same size from three branches are concatenated at the end of this module, producing an $(\frac{S_RW}{5}, 3)$ tensor.

\subsubsection{Deep Spatial-Temporal Residual Blocks}

Using deep residual blocks, we extract spatial-temporal features from the output of the multi-scale convolution. Each residual block contains five layers, two sets of 1D-convolutional layer and a batch normalization layer with a ReLU activation located in between, as illustrated in Figure \ref{fig:rrwavenet}. The input of each residual block is the concatenation between the input and the output from the previous block.

Eight residual blocks are placed one after another with increasing filter sizes. Instead of convolutional filters of size 1 as in the previous module, we increased the filter sizes, $F_i$, of the two 1D-convolutional layers every two residual blocks, which correspond to $F_1=F_2=64, F_3=F_4=128, F_5=F_6=256, F_7=F_8=512$. This results in the shape of the final tensor which is equal to $(\frac{S_RW}{5}, 3+2\sum_i{F_i})$.

\subsubsection{Respiratory Rate Estimator}
Placing at the beginning of the last module, a global average pooling layer converts the output from the residual blocks back to one dimensional of size $(3+2\sum_i{F_i})$. Three sets of an activation layer and a fully-connected layer then squeeze the input to the sizes of $128$, $64$, and $1$ which finally refers to RR, respectively.


\section{Experimental Setup}
\label{sec:experiments}   
In this section, we evaluate performance of \emph{RRWaveNet} using four datasets. In two experiments, we first compare performance of RRWaveNet to other state-of-the-art methods, including deep neural networks that have used either deep learning or hybrid approach, as well as the procedural signal processing approach on all four datasets. Following this, we apply transfer learning to estimate RR in {\color{red}other two datasets that were} obtained using a different device, WESAD and SensAI. We also compare performance when different source datasets are used for pretraining.

\subsection{Datasets}
\label{sec:dataset}

\begin{figure}[!t]
    \centering
  \includegraphics[width=\columnwidth]{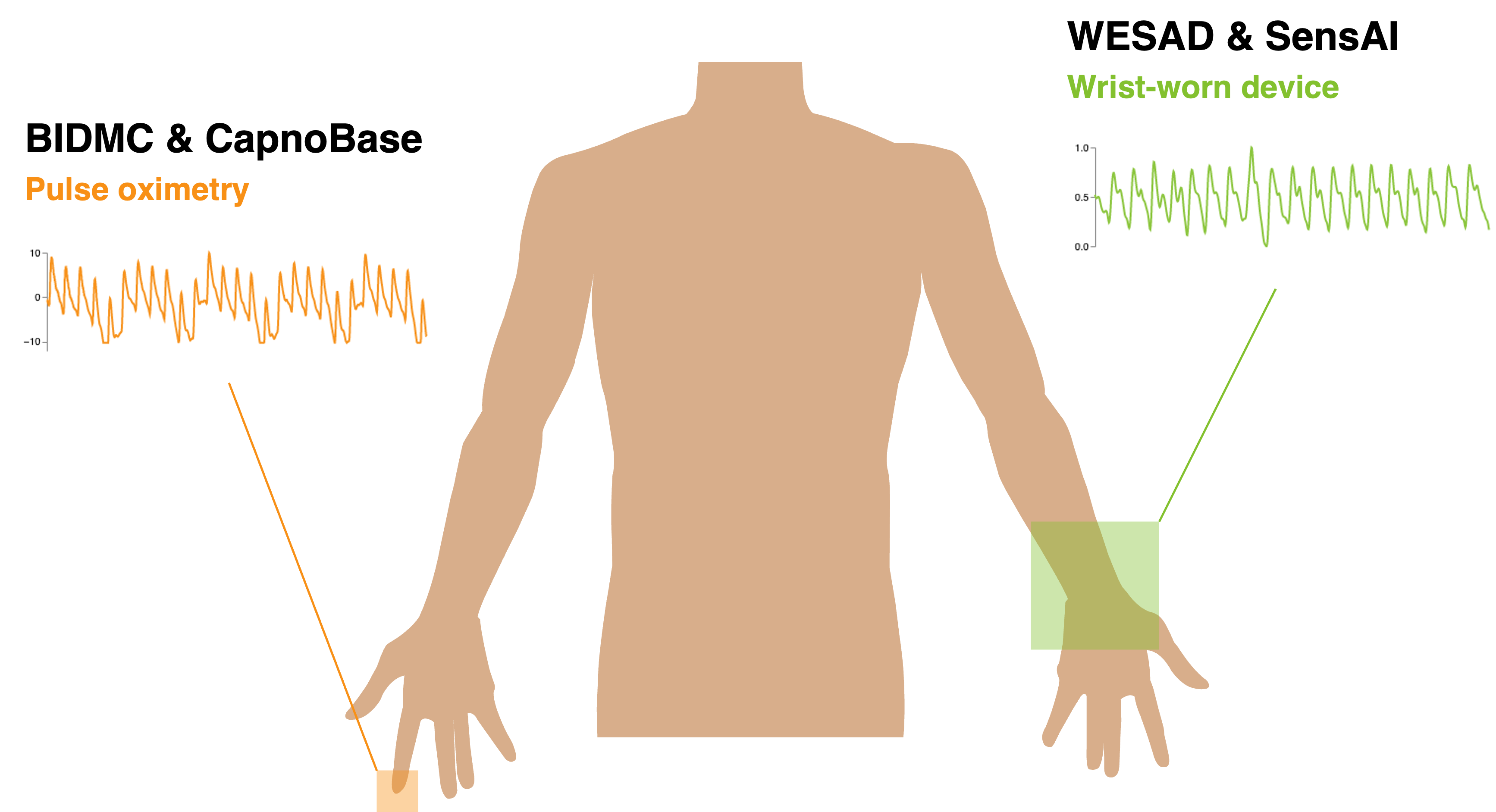}
  \caption{For performance evaluation, our study uses data from various data sources with different characteristics. WESAD and SensAI are noisy but large datasets obtained using a wrist-worn PPG sensor. In contrast, the BIDMC and CapnoBase datasets are higher quality datasets that used pulse oximeters attached to the fingertip.}
  \label{fig:body}
\end{figure}

To evaluate performance and test generalizability of the model, four benchmark datasets were used. These were obtained from various data sources, as shown in Figure \ref{fig:body}, which provided the different signal qualities and RR distributions.

\subsubsection{BIDMC \cite{Pimentel2017}}

The BIDMC dataset consists of electrocardiogram (ECG), pulse oximetry (PPG), and impedance pneumography respiratory signals acquired from intensive care unit (ICU) patients. The data was acquired by randomly selecting critically ill patients undergoing hospital care at the Beth Israel Deaconess Medical Centre (Boston, MA, USA). Two trained annotators manually annotated individual breaths in each recording using the respiratory impedance signal. The dataset contains 8-minute recordings of ECG, PPG, and impedance pneumography signals from 52 adult patients aged from 19 to more than 90, including 32 females.

\subsubsection{CapnoBase \cite{capnobase_paper}}
The CapnoBase dataset comprises of 8-minute recordings of electrocardiogram (ECG) and pulse oximetry (photoplethysmogram, PPG) along with capnometry signals from 42 subjects (13 adults, 29 children and neonates). The signals were acquired during elective surgery and routine anesthesia. The dataset also includes the inhaled and exhaled carbon-dioxide (\ch{CO_2}) signal labelled by the research assistance \cite{capnobase_paper} that can be utilized as the reference breathing rate.

\subsubsection{WESAD \cite{wesad}}
The multimodal WESAD dataset includes physiological and mobility data from wrist-worn (Empatica E4) and chest-worn (RespiBAN) devices. Data were acquired from 15 subjects, and contains multiple features, pulse rate, ECG, body temperature, as extracted from a wrist worn device and blood volume pulse (BVP) and respiration extracted by the chest-worn devices.

{\color{red}
\subsubsection{SensAI\protect\footnote{SensAI's official website: https://sensailab.com}}
Named after the wearable wrist-worn device SensAI, the dataset was collected from 50 Vidyasirimedhi Institute of Science \& Technology (VISTEC) staff and students who participated in the napping experiment. The dataset includes PPG signals, obtained from the SensAI device, respiratory inductance plethysmography (RIP), collected from a research grade device BiosignalsPlux, along with biosignals from other devices. 
}

\subsection{Training Strategy}
Since RR estimation is a regression task, we use mean-squared error (MSE) between the prediction and ground truth RR as the loss function, as this prioritizes penalization of high-error predictions. AdaBelief \cite{adabelief}  is used as the optimizer, which combines the strengths of both the Adam and the SGD optimizer -- adaptivity and generalization. We apply Hyperband\cite{hyperband} to tune the hyperparameters on the combination of the datasets. It offers the optimal number of residual layers and their number of features for our architecture, as well as a learning rate of $10^{-4}$ and an epsilon of $10^{-13}$ for the optimizer. We trained the model with $69420$ epochs to ensure convergence. However, to avoid overfitting that may occur in deep neural networks, we applied an early stopping technique with a patience of $5$ epochs based on the validation loss. This typically terminates the training within hundreds of epochs. We also decrease the learning rate when the validation loss has not decreased for $4$ consecutive epochs by a factor of at least $0.25$.
\newline

\begin{figure}[!t]
  \includegraphics[width=\columnwidth]{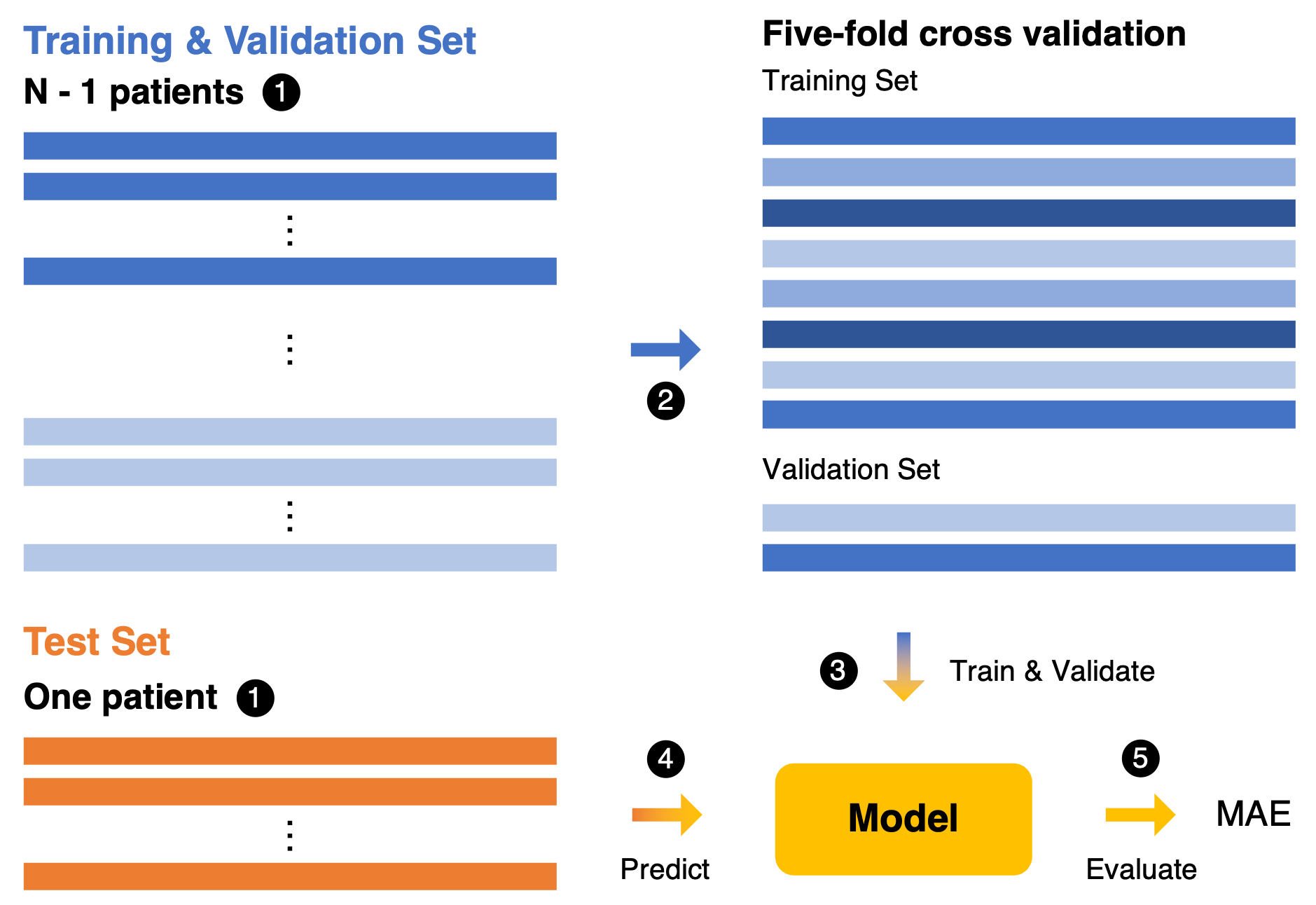}
  \caption{Our evaluation method as used in both experiments is based on the leave-one-out validation method which ensures subject independence. All patient samples, labelled with distinct colors, are one-dimensional with length of $S_RW$. As sequentially labelled using numbers above, first, samples from one patient are left as the test set, while those from the remaining patients belong to the training and validation set. The latter is further divided with a ratio of 4:1 to train the model with a five-fold cross validation. Each fold evaluates the test samples and outputs one MAE, which is averaged over the five folds. Each patient takes turn as the test set.}
  \label{fig:evaluation}
\end{figure}

\subsection{Experimental Evaluation}

In both experiments, we use the mean absolute error (MAE) in breaths per minute as the evaluation metric. Defined as shown in Equation \ref{eq:mae}  below, MAE is the sum of the absolute difference between the ground truth RR, $y_i$, and the predicted RR from the model, $y'_i$, and averaged over all $n$ testing samples.

\begin{align}
    \text{MAE} = \frac{60}{nW}\sum_{i=1}^{\color{red}n} |y_i-y'_i|
    \label{eq:mae}
\end{align}

We take into account the bias towards specific patients by performing a leave-one-out validation. As illustrated in Figure \ref{fig:evaluation}, each approach is evaluated subject-independently, thus exactly once for each patient by leaving the samples of the particular patient for later evaluation and training the model using the samples from the remaining patients. Applying the leave-one-out evaluation, all of the results reported in Table \ref{table:1} and Table \ref{table:2} indicate the average MAE and the standard deviation across all patients. However, since \cite{Pimentel2017} relies on purely signal processing, each PPG signal is processed and evaluated individually.

\subsection{Experiment I: Performance Comparison}
\label{sec:exp1}



\begin{table}[!t]
\caption{The comparison of respiratory rate estimators: \\
Mean absolute error (MAE) presented with mean value and standard deviation (SD)}
\centering
\resizebox{\columnwidth}{!}{
\begin{tabular}{llccc}

\toprule[0.2em]
\multirow{2}{*}{Dataset} & \multirow{2}{*}{Model} & 
\multicolumn{3}{c}{Window Size (seconds)} \\
\cmidrule[0.1em]{3-5}
& & $W=16$ & $W=32$ & $W=64$ \\ 

\midrule[0.1em]

\multirow{6}{*}{\emph{BIDMC}} & \textbf{RRWaveNet} & \textbf{1.87 $\pm$ 0.95} & \textbf{1.62 $\pm$ 0.86} & \textbf{1.66 $\pm$ 1.01} \\ 
& RespWatch & 1.88 $\pm$ 1.26 & 1.96 $\pm$ 1.99 & 1.66 $\pm$ 2.80 \\ 
& Bian et al. & 2.25 $\pm$ 1.26 & 2.46 $\pm$ 1.73 & 2.16 $\pm$ 1.23 \\
& RespNet & 2.45 $\pm$ 0.69 & 2.07 $\pm$ 0.98 & 2.06 $\pm$ 1.25 \\ 
& {\color{red}CycleGAN}* & {\color{red}-} & {\color{red}1.9 $\pm$ 0.3} & {\color{red}-} \\
& Autoregressive \cite{Pimentel2017} & - & 4 & 2.7 \\ 
& Reproduction of \cite{Pimentel2017} & 7.77 $\pm$ 5.98 & 8.97 $\pm$ 6.48 & 10.20 $\pm$ 6.67 \\

\midrule[0.1em]

\multirow{6}{*}{\emph{CapnoBase}} & \textbf{RRWaveNet} & \textbf{1.79 $\pm$ 1.37} & \textbf{1.86 $\pm$ 1.54} & \textbf{1.59 $\pm$ 1.08} \\
& RespWatch & 1.87 $\pm$ 1.17 & 2.09 $\pm$ 3.21 & 1.82 $\pm$ 3.15\\
& Bian et al. & 2.62 $\pm$ 3.27 & 2.36 $\pm$ 2.18 & 2.29 $\pm$ 2.29 \\
& RespNet & 5.40 $\pm$ 4.29 & 4.86 $\pm$ 4.79 & 4.48 $\pm$ 5.14 \\ 
& Autoregressive \cite{Pimentel2017} & - & \textbf{1.5} & 1.9 \\
& Reproduction of \cite{Pimentel2017} & 4.40 $\pm$ 6.54 & 8.26 $\pm$ 8.19 & 6.87 $\pm$ 8.35 \\
\midrule[0.1em]

\multirow{4}{*}{\emph{WESAD}} & \textbf{RRWaveNet} & \textbf{2.77 $\pm$ 0.62} & \textbf{2.19 $\pm$ 0.91} & \textbf{1.92 $\pm$ 0.96} \\
& RespWatch & 2.96 $\pm$ 0.47 & 2.46 $\pm$ 0.58 & 2.26 $\pm$ 0.65 \\
& Bian et al. & 3.36 $\pm$ 0.40 & 2.80 $\pm$ 0.36 & 2.45 $\pm$ 0.52 \\
& Reproduction of \cite{Pimentel2017} & 9.38 $\pm$ 5.25 & 11.34 $\pm$ 5.40 & 12.95 $\pm$ 5.15 \\

\midrule[0.1em]

\multirow{4}{*}{\emph{SensAI}} & \textbf{RRWaveNet} & \textbf{2.49 $\pm$ 0.86} & \textbf{1.83 $\pm$ 1.12} & \textbf{1.23 $\pm$ 0.61} \\
& RespWatch & 2.52 $\pm$ 0.78 & 1.95 $\pm$ 1.15 & 1.45 $\pm$ 0.65 \\
& Bian et al. & 2.54 $\pm$ 1.07 & 2.07 $\pm$ 1.37 & 1.59 $\pm$ 0.86 \\
& Reproduction of \cite{Pimentel2017} & 12.96 $\pm$ 4.86 & 14.93 $\pm$ 4.26 & 16.29 $\pm$ 3.74 \\

\bottomrule[0.2em]
\end{tabular}
}
{\justify {\color{red}* The MAE was taken directly from the paper \cite{Aqajari2021} which evaluated using $W=30$. \par}}
\label{table:1}
\end{table}

After data preprocessing, the performance of each RR estimation model is compared. \emph{RRWaveNet} and the other five baseline approaches presented in Table \ref{table:comparison} are evaluated on four benchmark datasets as described in Section \ref{sec:dataset} and across three window sizes: $16$, $32$, and $64$ seconds. 

Since RespNet is not an end-to-end model that directly predicts RR but reconstructs the \ch{CO_2} signal from the input PPG. RR was calculated by finding and counting the peaks in the predicted \ch{CO_2} signal. Peaks were detected by simply comparing the values between neighbouring points in the signal. Using this approach, a horizontal distance threshold needed to be set to prevent overwhelming false positives. For a fair comparison, we choose the best distance threshold based on the training set and applied it to the test set at evaluation.

{\color{red}
CycleGAN, too, reconstructs the respiratory signal from the input PPG and later uses an external library to capture the peaks from the generated respiratory signal. Unlike our leave-one-out evaluation, the authors evaluated their approach using subject-independent 5-fold cross validation. Their results on the BIDMC dataset are included in Table \ref{table:1} for comparison.
}

The results, with MAE as the evaluation metric, are compared in Table \ref{table:1}. \emph{RRWaveNet} outperformed all other deep learning networks across all datasets and window sizes with smaller MAEs and variances. Signal processing using multiple autoregressive models produces better results than our model only at $W=32$ for the CapnoBase dataset. Still, autoregressive models failed to perform well across patients and yield high variances between the predictions according to \cite{Pimentel2017}. 

Window size has been shown to be another important factor that can affect the MAE. Using the longest window size of $64$ seconds, all models output the least amount of error compared to other window sizes since it does not require the increased inference from shorter time periods to determine the RR in a 60-second interval (BPM). However, it is inconclusive to claim that using a longer window size always yields a better result, as it simply shows that errors and artifact have less effects when large amounts of data and low resolution is used.

\begin{table}[]
\color{red}
\caption{The Mean absolute error after applied transfer learning by fine tuning the WESAD and SensAI datasets by using the pre-trained weight collected the high quality ICU dataset}
\centering
\begin{tabular}{lllll}

\toprule[0.2em]

\multicolumn{1}{c}{\multirowcell{2}{Target\\Dataset}} & \multicolumn{1}{c}{\multirowcell{2}{Pretraining\\Dataset}} & \multicolumn{3}{c}{Window size (seconds)} \\ 

\cmidrule[0.1em]{3-5}

& \multicolumn{1}{c}{} & \multicolumn{1}{c}{$W = 16$} & \multicolumn{1}{c}{$W = 32$} & \multicolumn{1}{c}{$W = 64$} \\ 

\midrule[0.1em]

\multicolumn{1}{c}{\multirow{4}{*}{\emph{WESAD}}} & None & \multicolumn{1}{c}{\textbf{2.77 $\pm$ 0.62}} & \multicolumn{1}{c}{2.19  $\pm$ 0.91} & \multicolumn{1}{c}{1.92 $\pm$ 0.96} \\

\cmidrule[0.1em]{2-5}

& BIDMC & \multicolumn{1}{c}{2.81 $\pm$ 0.89} & \multicolumn{1}{c}{\textbf{1.87  $\pm$ 0.32}} & \multicolumn{1}{c}{\textbf{1.52 $\pm$ 0.50}} \\

& CapnoBase & \multicolumn{1}{c}{2.85  $\pm$ 0.29} & \multicolumn{1}{c}{2.07  $\pm$ 0.30} & \multicolumn{1}{c}{1.99  $\pm$ 0.57} \\ 


\midrule[0.1em]

\multicolumn{1}{c}{\multirow{4}{*}{\emph{SensAI}}} & None & \multicolumn{1}{c}{2.49 $\pm$ 0.86} & \multicolumn{1}{c}{1.83 $\pm$ 1.12} & \multicolumn{1}{c}{1.23 $\pm$ 0.61} \\

\cmidrule[0.1em]{2-5}

& BIDMC & \multicolumn{1}{c}{\textbf{2.30 $\pm$ 0.76}} & \multicolumn{1}{c}{\textbf{1.77 $\pm$ 1.02}} & \multicolumn{1}{c}{\textbf{1.07 $\pm$ 0.57}} \\

& CapnoBase & \multicolumn{1}{c}{2.37 $\pm$ 0.79} & \multicolumn{1}{c}{1.86 $\pm$ 1.26} & \multicolumn{1}{c}{1.23 $\pm$ 0.62} \\







\bottomrule[0.2em]

\end{tabular}
\label{table:2}
\end{table}















\subsection{Experiment II: Predicting RR Independent PPG Data from Wearable Device Using Transfer Learning}
\label{sec:exp2}
Our second experiment aimed to assess performance of our proposed model with PPG signals obtained from wearable devices. Different from the other two datasets, which were acquired from ICU patients via pulse oximeters, the WESAD {\color{red} and the SensAI} dataset {\color{red}were} collected from a wrist-worn device in a lab environment. In this case, we used transfer learning to take advantage of the high-quality ICU data so that it can be used with more confidence in a smaller dataset containing a signal of lower quality. 

We compare the performance of \emph{RRWaveNet} pretrained on one of the two high-quality ICU datasets: BIDMC and CapnoBase. The model is pretrained using the same training strategy adopted in the previous section. The pretrained weights are finally fine-tuned on the WESAD {\color{red} and the SensAI} dataset by extending the training with the same settings. This knowledge transferring strategy, as demonstrated and shown in Table \ref{table:2}, significantly reduces the error variance in almost all pretraining sets and window sizes. 

Using the BIDMC dataset as the source domain turns out to be most beneficial for WESAD predictions while the model does not perform well when the pretraining set is only the CapnoBase dataset via MAE. {\color{red}A similar trend can be seen when the target domain is the SensAI dataset, showing least MAEs when the BIDMC is utilized for pretraining.}

\section{Result and Discussion}
\label{sec:discussion}
{\color{red}This section is divided into five sections.} First, we emphasize the capability of \emph{RRWaveNet} which outperforms the baselines. {\color{red}Second, we highlight the importance of the multi-scale convolution module by conducting a simple ablation study.} Third, we interpret the understandability of \emph{RRWaveNet}'s effectiveness using explainable AI. Fourth, we discuss the benefits of using transfer learning on wearable datasets. Finally, we propose a few possibilities of employing \emph{RRWaveNet} in clinical applications.

\begin{figure*}[!t]
  \includegraphics[width=0.313\textwidth]{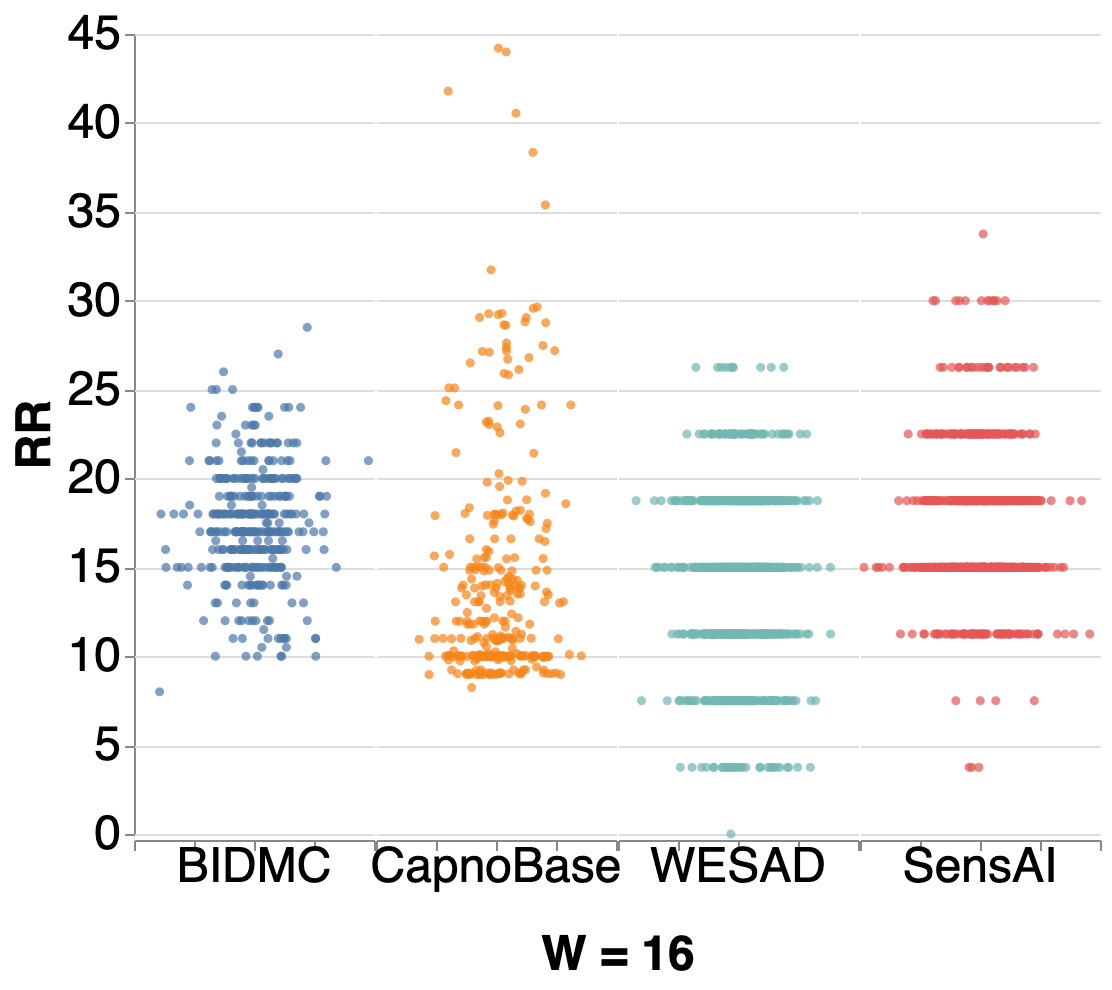}\hfill
  \includegraphics[width=0.3\textwidth]{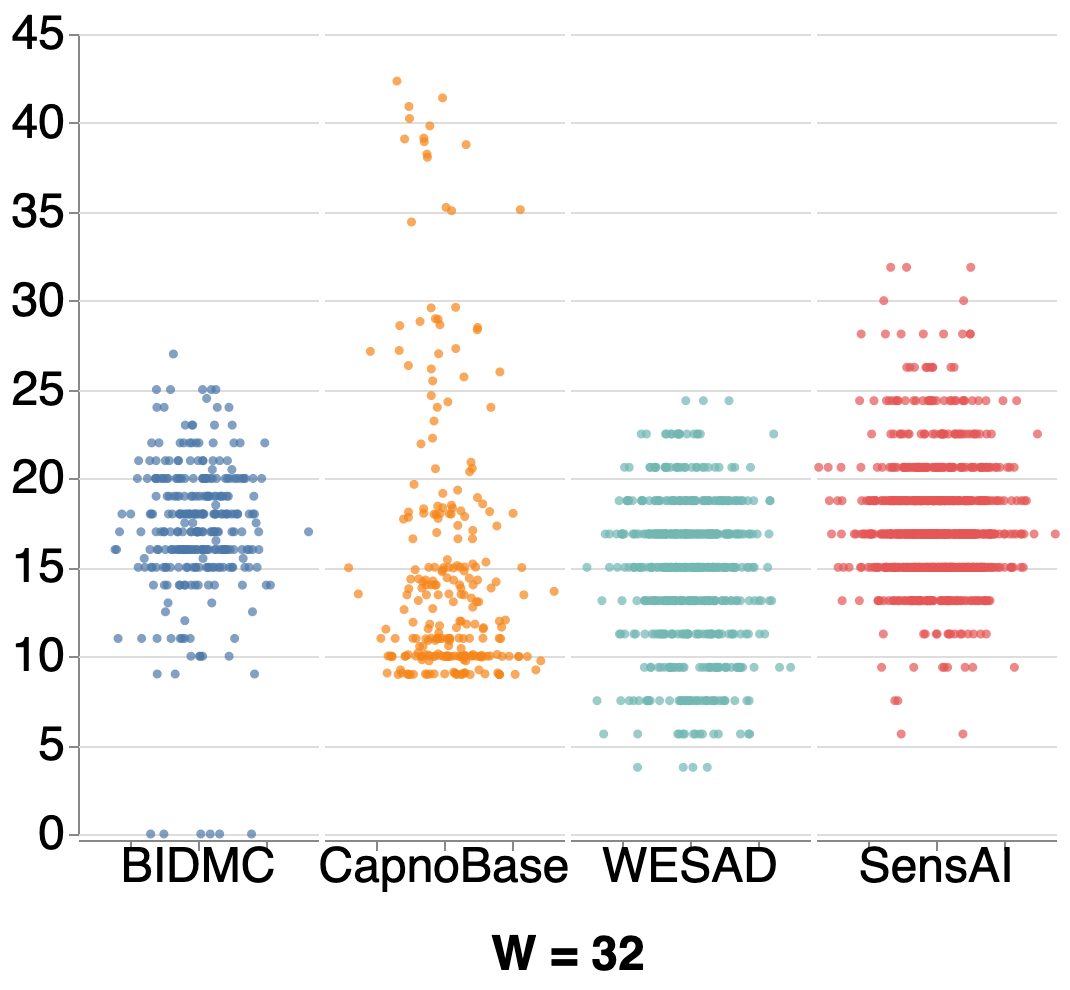}\hfill
  \includegraphics[width=0.3\textwidth]{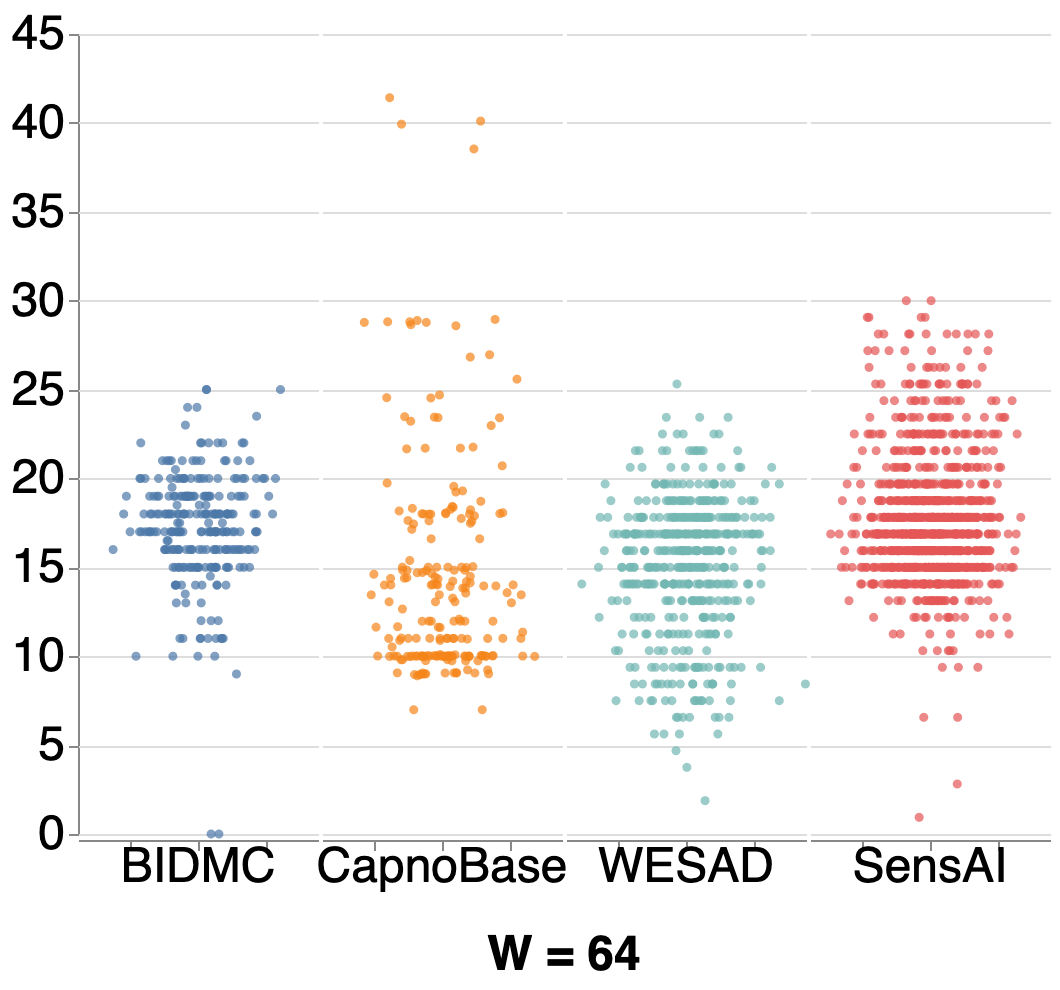}
  \caption{The distribution of respiratory rate containing 4000 samples from each dataset. The CapnoBase dataset includes respiratory rates that are higher than 35 BPM, which do not appear in other three datasets.}
  \label{fig:rr_dist}
\end{figure*}

\begin{figure}[!t]
  \includegraphics[width=\columnwidth]{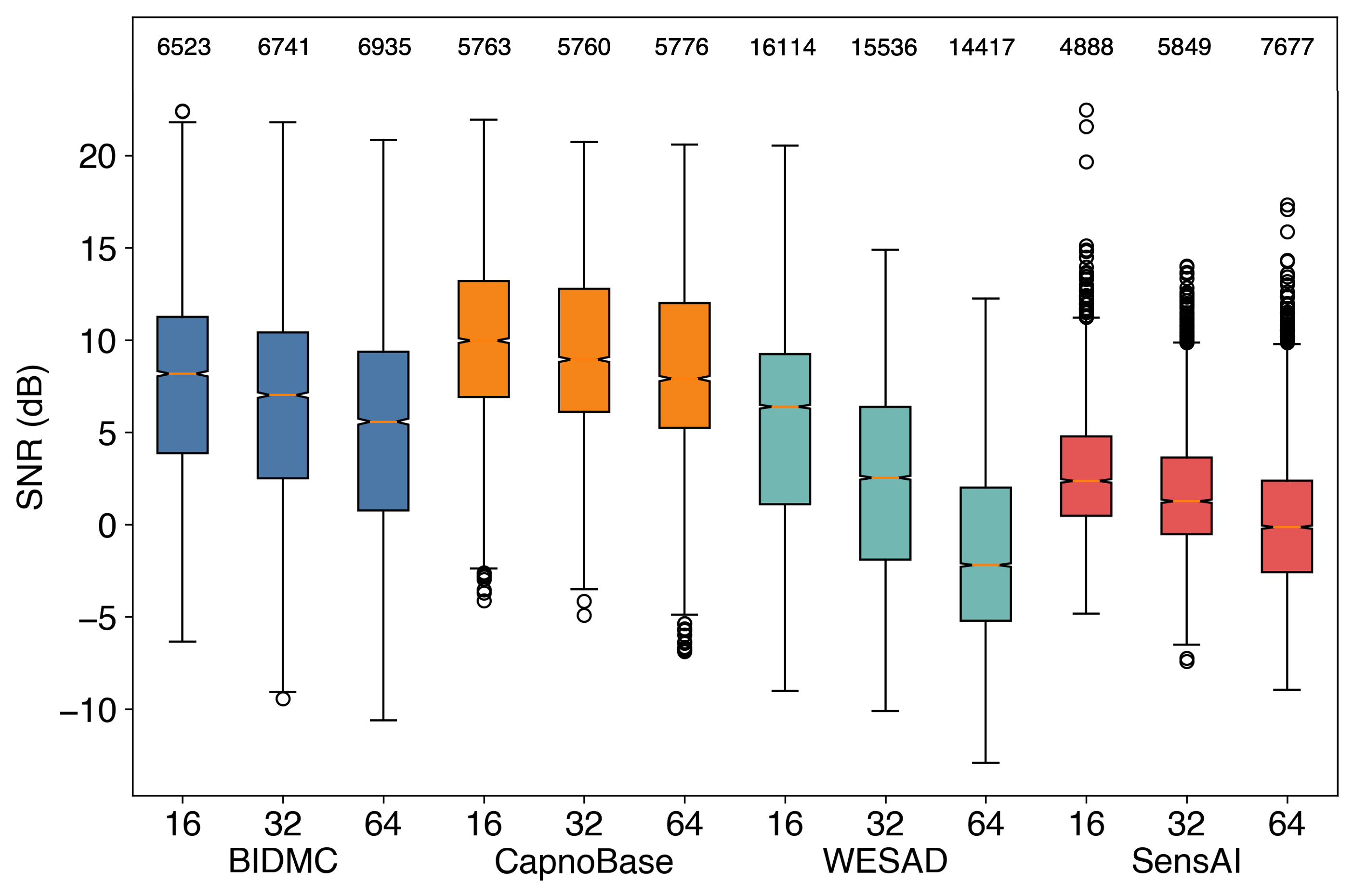}
  \caption{Signal-to-noise ratios (SNR) indicate the quality of the data from each baseline dataset in this study. High SNR value indicates better signal quality, and vice versa. The numbers of samples are shown on top of the boxes.}
  \label{fig:snr}
\end{figure}

\subsection{Performance Analysis with Other Baselines}


The subject-independent performance results in Table \ref{table:1} show that \emph{RRWaveNet} outperforms other state-of-the-art across different window sizes and datasets. This study focuses on the model's simplicity since the ultimate goal is to create a model which can perform on a wearable device with low processing power and allows continuous monitoring. \emph{RRWaveNet} has an advantage over the autoregressive approach \cite{Pimentel2017}. As it is an end-to-end architecture, it does not require handcrafted feature engineering and generates a robust performance in different samples and datasets using learnable parameters. Among the remaining deep learning-based models, not only it can produce the lowest error across all datasets and window sizes, but also \emph{RRWaveNet} can estimate RR as the output from raw PPG (end-to-end architecture). It requires a lower sampling rate for the input signal (50 Hz for \emph{RRWaveNet} instead of 256 Hz) as shown in Table \ref{table:comparison}. This flexible requirement proves the feasibility of working on low processing devices as we expected.

{\color{red}
\subsection{An Ablation Study on Multi-Scale Convolution Module}
To signify the impact of the multi-scale convolution module of \emph{RRWaveNet}, we conducted an ablation study on a version of \emph{RRWaveNet} without the multi-scale convolution module (the leftmost component of Figure \ref{fig:rrwavenet}), which we will refer to as \emph{RRWaveNet-Plain} from this point forward. We re-trained \emph{RRWaveNet-Plain} using a sample window size of 16 seconds on all four datasets, while ensuring that all training strategies remained the same in order to maintain a fair comparison. 

\begin{table}[t!]
    \caption{Ablation study between RRWaveNet and RRWaveNet-Plain\\Mean absolute error (MAE) presented with mean value and standard deviation (SD)}
    \centering
    \resizebox{\columnwidth}{!}{%
    \begin{tabular}{l@{\hspace{4em}}l@{\hspace{6em}}l}
    
    \toprule[0.2em]
    
    \textbf{Dataset} & \textbf{RRWaveNet} & \textbf{RRWaveNet-Plain} \\ 
    
    \midrule[0.1em]

    \emph{BIDMC}     & \textbf{1.89 $\pm$ 0.95} & 1.93 $\pm$ 1.01 \\
    \emph{CapnoBase} & \textbf{1.79 $\pm$ 1.37} & 2.35 $\pm$ 2.67 \\
    \emph{WESAD}     & \textbf{2.77 $\pm$ 0.62} & 2.97 $\pm$ 0.32 \\
    \emph{SensAI}    & \textbf{2.49 $\pm$ 0.86} & 2.58 $\pm$ 0.84 \\
    
    \bottomrule[0.2em]
    \end{tabular}}
    \label{table:rrwavenetplain}
\end{table}

Table \ref{table:rrwavenetplain} suggested that \emph{RRWaveNet} indeed outperformed \emph{RRWaveNet-Plain} when considering the mean MAE of each experiment. 
These results provide a strong evidence for the effectiveness of the multi-scale convolution module contributing to the performance of \emph{RRWaveNet}.
}

\subsection{Performance Improvement on the {\color{red}Wrist-Worn Datasets} Using Transfer Learning}

As stated in the introduction, we want a robust model to perform consistently well across datasets. We would like to use the benefits of transfer learning to address this issue, {\color{red}especially when the target dataset has lower quality}. Table \ref{table:2} shows that transfer learning enhances the estimation quality compared to training the model from scratch. 

Compared to BIDMC, using CapnoBase as the pretraining dataset did not yield results as promising as BIDMC provided. An explanation can be inferred from Figure \ref{fig:rr_dist}, showing the distribution of respiratory rates in each dataset and each window size. While BIDMC, WESAD, {\color{red}and SensAI} datasets were collected from older subjects, the CapnoBase dataset includes the samples acquired from children and neonates with typically higher BPM compared to adults. This results in the disparity between the distribution found in the CapnoBase dataset and those from the others, which becomes challenging for the model to transfer the acquired knowledge from one domain to another. 

Possible improvements could further increase the model's performance and improve the domain adaptability of the model. As the results indicate the likely poor performance if the test samples are out-of-distribution, applying the transfer learning potentially offers better performance if another dataset that covers the covariance shift presents. However, training a robust model that performs well in multiple distributions must take into account the domain shifts. Possible workarounds may include continual learning and finding a shared representation between the distributions.

\subsection{Effects of Dataset Characteristics to RR Estimation}
{\color{red}
   To support our experimental setup of transfer learning, which we handpicked domain and target datasets as above, we compare one of the dataset characteristics, namely the signal quality of the four datasets. This might also align well with the MAEs evaluated on each dataset despite the choice of RR estimating model. 
   
   Two different types of datasets were used throughout this study: the clinical datasets (CapnoBase and BIDMC) and the wearable datasets (WESAD and SensAI). 
    Following a recent study on quantifying the signal quality of rPPG signal \cite{Li2022snr}, we utilized SNR, or signal-to-noise ratio, as a measure of the data quality to illustrate the dataset characteristics. The SNR of a signal window was obtained by first converting the signal to the frequency domain using FFT. The final value (in dB) followed the formula below
    \begin{align}
        \text{SNR}_{\text{dB}} = 10\log_{10}\frac{P_{\text{signal}}}{P_{\text{noise}}}
    \end{align}
    where $P_{\text{signal}}$ was extracted from the window's peak frequency and its first harmonics and the remaining frequencies were accounted to $P_{\text{noise}}$.
    
    We calculated SNR of each window from all datasets and visualized in Figure \ref{fig:snr}. Clinical datasets revealed higher SNRs for all window sizes, indicating better data quality and thus yielding smaller MAEs. Similarly, wearable datasets provided relatively poor SNR values, correlating to greater MAEs. These observations showed a strong correlation between the data quality and efficiency of RR estimation. They also inspired choosing high-quality clinical datasets as the pretraining datasets to improve the prediction performance when the low-quality target dataset was collected from a wearable device.
}

\begin{figure}[!t]
  \includegraphics[width=\columnwidth]{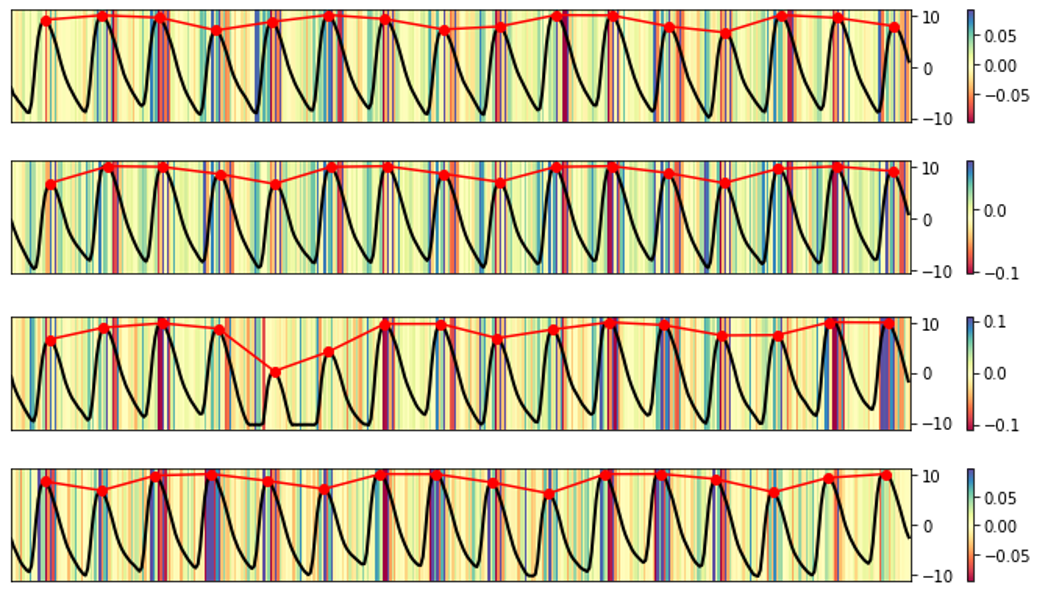}
  \caption{The illustration of SHAP scores on several 4-second segments from the CapnoBase dataset. The background colormap denotes the relative SHAP score computed for each segment. The red signals denotes the respiratory-induced intensity variation (RIIV) of the PPG signal colored in black.}
  \label{fig:shap}
\end{figure}

\subsection{Explanation of \emph{RRWaveNet}'s Effectiveness with SHAP}

While the MAE evaluation metric provides an indication of \emph{RRWaveNet}'s ability to accurately estimate RR, it does not fully reveal the reasoning behind each RR estimation. To gain a deeper understanding of our model, it is of interest to us to explain which features in the input exhibit importance towards the model's prediction. The SHapley Additive exPlanations (SHAP) \cite{shap} is particularly useful in this regard, as it allows us to interpret how much each input feature contributes to the prediction. SHAP calculates feature attribution by excluding some input features and measuring the impact on the prediction. Features that have a significant effect on the prediction will result in a higher SHAP score. In our case, we applied SHAP to identify which sections of the PPG signal are most important for \emph{RRWaveNet}'s estimation of the RR.

The implementation of SHAP explanations for \emph{RRWaveNet} utilizes connections with DeepLIFT, known as Deep SHAP (available in the official SHAP package) \cite{shap}. Deep SHAP calculates the feature importance $\Delta t$ with respect to a reference value, which in this case is a sample from the training set, as shown in Equation \ref{eq:deeplift}.
\begin{align}
    \Delta t = \sum_{i=1}^{N}C_{\Delta x_{i} \Delta t}
    \label{eq:deeplift}
\end{align}

For a targeted value $t$, let $x_1, x_2, \dots, x_n$ be the values assigned to artificial neurons connected to $t$. We calculate $C_{\Delta x_{i} \Delta t}$ by computing the linear difference between the reference value and the value at $x_i$, which allows us to quantify the magnitude and direction of change of $t$ induced at that neuron \cite{deeplift}. All non-linear components of \emph{RRWaveNet} are linearized using the SHAP equations.

{
\color{red}
Figure \ref{fig:shap} shows the SHAP values for PPG segments from the CapnoBase dataset, along with annotations of the detected PPG peaks. It can be seen that higher SHAP scores generally coincide with the detected PPG peaks, indicating that peak heights and positions are taken into consideration in the RR estimation process. In the context of RR estimation from PPG, these peaks are important in determining the Respiratory-induced Variation (RIXV) signals derived from the PPG, such as RIIV (intensity), RIAV (amplitude), and RIFV (frequency). These variation signals are caused directly by respiratory activities and are emphasized for their importance in RR estimation \cite{Pimentel2017}. As an example, we visualized the RIIV of the PPG signal, as shown by the red-lined signal in Figure \ref{fig:shap}. In our illustration, the high SHAP scores correlate with the peaks of RIIV, which suggests that the change in blood intensity caused by changes in oxygen saturation due to respiratory activities is taken into consideration in the RR estimation process.

The claim can be further supported by a study conducted by Yasuma et al.
This study has pointed out the intertwining correlations between the circulatory and respiratory systems from the common neurological control of the autonomic nervous system \cite{yasuma2004respiratory}. For example, certain activities in the cardiac cycle may influence diaphragmatic function. As a result, aspects of respiratory activities are reflected positions and variability of EKG R-peaks peaks. As these features from the EKG are proven correspond with those from the PPG \cite{lin2014comparison}, a conclusion can be drawn that PPG peaks positions are physically correlated to respiration.

In summary, the results suggest \emph{RRWaveNet}'s ability to selectively detect PPG peaks, which is an important feature used in RR estimation. The implication also extends to other PPG-to-vital sign estimation tasks that require PPG peak detection in the future.}

\subsection{Clinical Applications}

The promising results from the tables reveal possible improvements in hospital healthcare using technological aids. Three examples are listed below.

\subsubsection{RR Measurement in Out-Ward Patients} 

Section \ref{sec:experiments} has shown the practicality of precise and instant RR estimation of a remote patient without the necessity of model fine-tuning. The results from both experiments have shown that an accurate RR can be extracted from the PPG signal collected using wrist-worn devices instead of through medical appointments in well-equipped facilities.

\subsubsection{Automated Continuous Monitoring of In-Ward Patients}

For conditions that are highly sensitive to respiration e.g. sepsis, respiratory distress, and pneumonia, continuous RR estimation becomes very crucial. The capability of continuous RR measure tends to significantly reduce the workload of the medical staff along with the replacement of an automatic alerting system when the estimated RR is out of the normal range.

Regarding this application, we provide a brief analysis of the feasibility of employing this application in the hospital ward through a specific scenario. Motivated by the early warning score (EWS), which has been effectively used to recognize deterioration in patients \cite{Smith_2013, Alam_2014, Shamout_2020} as well as to pre-screen sepsis which is specifically developed in \cite{Suttapanit_2022}, we adopt its RR scoring rubrics as our reference scale with the intention to detect the abnormality of RR and alert the medical staff accordingly. According to \cite{Smith_2013}, RR in each predetermined range is assigned to a score between 0 (normal) and 3 (most severe).

\begin{figure}[!t]
  \includegraphics[width=\columnwidth]{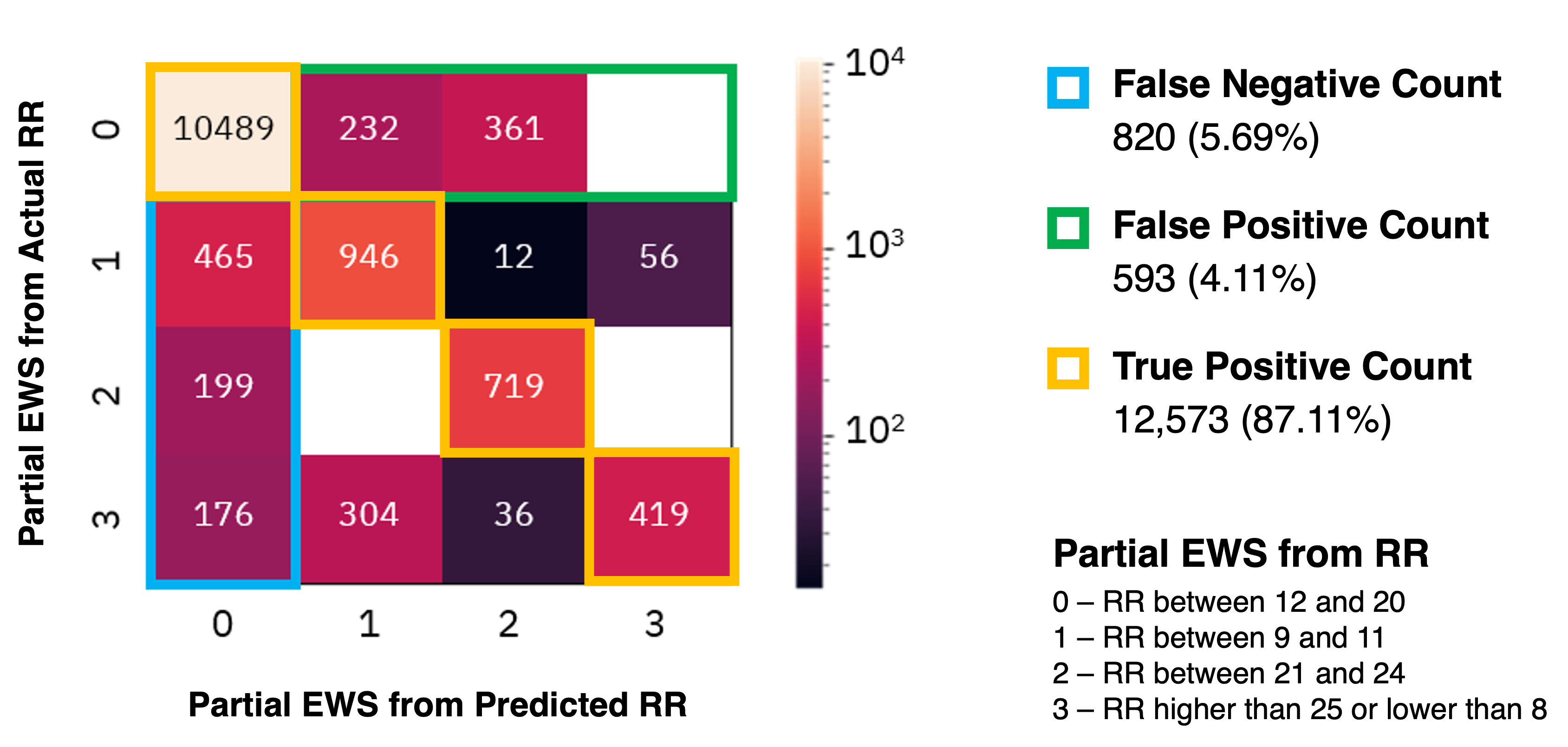}
  \caption{The confusion matrix (left) shows the classification accuracy of a specific scenario when RR is categorized into four labels (0 to 3) according to early warning score (EWS) assessment \cite{Smith_2013}. It depicts the performance of \emph{RRWaveNet} with weights pretrained from the BIDMC dataset, predicting RR on 64-second windows. The metrics (right) indicate the feasibility of applying our approach for an automated continuous monitoring using EWS rubric as a reference.}
  \label{fig:cm}
\end{figure}

To emphasize the possibility of using wrist-worn devices as the signal collector, we choose ground truths and \emph{RRWaveNet}'s predictions on 64-second windows from the WESAD dataset and evaluate them on the EWS's RR rubrics. The confusion matrix in Figure \ref{fig:cm} shows the classification accuracy of \emph{RRWaveNet}, with an F1 score of $71.76\%$, when its RR estimation is converted to partial EWS. Another important metric that must be mentioned is the false negative rate (FNR) which we need to minimize to prevent losses. Our best model achieves $5.69\%$ FNR although it is not designed to predict partial EWS regarding the RR. On the other hand, a false positive rate (FPR) of $4.11\%$ implies the number of false alarms that may occur in hospital wards and draws immoderate attention from the medical staff. However, as the purpose of this evaluation is to suggest a possibility from a practical perspective, the model must be heavily repurposed and refined to eliminate the false negatives before adopting the RR EWS estimator in real clinical practices.


\subsubsection{Other Related Applications}
Other than the RR estimation task, a possible modification can be made to the model architecture to estimate the heart rate and other vital signals, given the PPG signal as in this study. Utilizing biosignals other than PPG to perform the task can be successful with proper modifications as well. 

Furthermore, we can extend the capability of the model to predict a high-level indicator such as EWS, directly aiming to implement an automatic alert for in-ward monitoring. A reliable framework would potentially reduce the workload of the technician from routine patient assessment.




\section{Conclusion}
\label{sec:conclusion}
\emph{RRWaveNet}, which is a compact end-to-end deep learning architecture specialized for respiratory rate estimation from PPG signal, is proposed in this study. It consists of a multi-scale convolution network, deep spatial-temporal residual blocks, and an estimator module, which captures the necessary features from the raw PPG input signal and learns to predict the respiratory rate at any given window size. We evaluated its performance on four datasets in comparison with other state-of-the-art, both signal processing-based and deep learning-based, exhibiting the outperforming results from \emph{RRWaveNet} over other baselines. Transfer learning is also introduced to tackle the scarcity problem and the low quality of the data, which turns out to be strongly effective, especially in the WESAD dataset, but specific to the data distribution. This offers the possibility of using our model to develop applications for widely available wearable devices.

\section{Acknowledgement}
\label{sec:acknowledgement}

We would like to show our gratitude to Theerach Temiyasathit for sharing his insights with us during the writing of this manuscript. 


\appendices


\bibliographystyle{IEEEtran}
\bibliography{citelist}

\end{document}